# Tuning the electronic structures and transport properties of zigzag blue phosphorene nanoribbons

Yipeng An,*,[†,‡] Songqiang Sun,[†] Mengjun Zhang,[†] Jutao Jiao,[†] Dapeng Wu,[§] Tianxing Wang [‡] and Kun Wang[∥]

[†]College of Physics and Materials Science & International United Henan Key Laboratory of Boron Chemistry and Advanced Energy Materials, Henan Normal University, Xinxiang 453007, China
E-mail: ypan@htu.edu.cn
[‡]Department of Physics and Astronomy, University of California, Irvine, California 92697, USA
[§]School of Chemistry and Chemical Engineering & International United Henan Key Laboratory of Boron Chemistry and Advanced Energy Materials, Henan Normal University, Xinxiang 453007, China
[∥]Department of Mechanical Engineering, University of Michigan, Ann Arbor, Michigan 48109, USA

*Abstract*—In recent years, single element two-dimensional (2D) atom crystal materials have aroused extensive interest in many applications. Blue phosphorus, successfully synthesized on Au (111) substrate by molecular beam epitaxy not long ago, shows unusual geometrical and electronic structures. We investigate the electronic structures and transport properties of zigzag blue phosphorene nanoribbons (ZBPNRs) by using a first-principles method, which can be obviously tuned *via* different groups (*i.e.*, -H, -O, and -OH) passivation on the both edges. The ZBPNRs-H and ZBPNRs-OH present a wide gap semiconductor property. While the ZBPNRs-O are metallic. Interestingly, the current–voltage (*I-V*) curves of ZBPNRs-O show a negative differential resistive (NDR) effect, which is independent on the ribbon width. The electric current through the ZBPNRs-O is mainly flowing along the both outside zigzag phosphorus chains *via* the way of P-P bond current. Through modifying the both edges with various functional groups, the ZBPNRs can display some important functional characteristics and become a candidate of NDR devices.

## I. INTRODUCTION

OVER the past decades, some two-dimensional materials, such as graphene [1]–[6], hexagonal boron nitride (h-BN) [7]–[9], silicon carbide (SiC) [10], [11], transition-metal dichalcogenides [12]–[15], phosphorene [16]–[18], and phosphorene-like structures [19], [20], have caused extensive interest, not only from a fundamental point of view to understand their intrinsic properties but also from a practical point of view to study their potential applications in future nano-devices. Phosphorene has a variety of allotropes based on their different atomic structure arrangement. It includes the common white, red, violet, and black allotropes, with the color defined by the bandgap. Phosphorene is studied with the hope of overcoming graphene's deficiencies, such as its zero bandgap. It has been emerged as a promising new two-dimensional material due to its intrinsic and tunable bandgap, high carrier mobility and remarkable in-plane anisotropic electrical, optical and phonon properties [21], [22]. With its excellent transport properties, phosphorene has attracted a remarkable interest for potential applications in high-performance thin-film electronics, mid- and near-infrared optoelectronics, and for the development of conceptually novel devices that utilize its anisotropic properties [23], [24]. For instance, the Zhang [16] group synthesized a few layers of black phosphorus in an experiment using the scotch tape-based mechanical exfoliation method. They peeled thin flakes from bulk crystals and fabricated the field-effect transistors based on few layer black phosphorus crystals with a thickness down to a few nanometers. Due to the strong anisotropic atomic structure of black phosphorus, its electronic conductivity and optical response are sensitive to the magnitude and the orientation of the applied strain, which appears as a promising method to design novel photovoltaic devices that capture a broad range of solar spectrum [25]. Xie *et al*. [26] investigated the electronic and transport properties of zigzag phosphorene nanoribbons (ZPNRs) by performing first-principles calculations. It presents a negative differential resistance behavior, and the peak-to-valley current ratio is up to 100 under low biases. Zhu *et al*. [27] first predicted the existence of a previously unknown phase of phosphorus, referred to as blue phosphorus. The blue phosphorus is hexagonal lattice crystals with a small out-plane distance of two sublattice sites, which is nearly as stable as black phosphorus. Lately, Chen *et al*. [28] synthesized single layer blue phosphorus on Au (111) substrate by molecular beam epitaxy for the first time. And the first-principles study shows that the nanoribbon edges will be significantly changed under strain which results in a linear decrease of the gap of blue phosphorus nanoribbons with deformation aggravation [29].

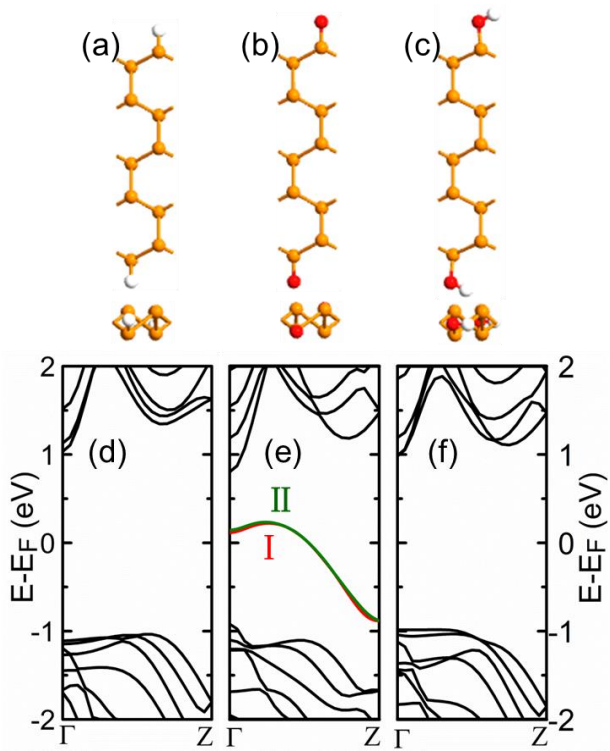

Fig. 1.    The geometric and electronic structures of ZBPNRs. (a) - (c) Top and side views of the 5-ZBPNR-H, 5-ZBPNR-O, and 5-ZBPNR-OH unit cell, respectively. (d) - (f) Band structures of the 5-ZBPNR-H, 5-ZBPNR-O, and 5-ZBPNR-OH, respectively.

In this work, we investigate the electronic structures and transport properties of zigzag blue phosphorene nanoribbons by a first-principles method, which can be obviously tuned through three different functional groups (*i.e.*, -H, -O, and -OH) passivation on both edges. It makes them display semi-conductive and metallic characteristics, respectively. This edge-passivation way makes the blue phosphorous nanoribbons present more abundant electronic transport phenomena and can be useful for the design of phosphorous-based nano-electronic devices.

## II. COMPUTATIONAL METHODS

We study the electronic transport properties of the ZBPNRs-based devices by density functional theory combined with the non-equilibrium Green's functions method (NEGF-DFT) [30]–[33]. The GGA-PBE is adopted as the exchange correlation functional [34], [35]. The wave-functions of valence electrons are expanded in double-zeta plus polarization (DZP) basis set. The Hamiltonian and electronic densities are calculated in a real space grid defined with a plane-wave cut-off 150 Ry. The atomic structures are optimized until the absolute value of force acting on each atom is less than 0.01 eV/Å. The $k$-points grid $1 \times 1 \times 100$ is used to sample the Brillouin zone of electrodes in the X, Y, and Z directions (transport direction), respectively.

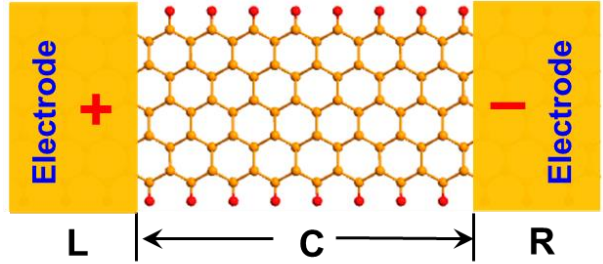

Fig. 2. The two-probe device model for the zigzag blue phosphorene nanoribbons.

## III. RESULTS AND DISCUSSION

Up on structural optimization of the unpassivated ZBPNRs, the interior part of the nanoribbons experiences negligible structural change, but the edges of the nanoribbons show some degree of deformations [26]. They are not stable and easily twisty like graphene nanoribbons, due to the strong dangling bond at the both edges. Therefore, we try to stabilize the ZBPNRs by passivating the both edges with -H, -O, and -OH groups. Fig. 1(a)-(c) shows the unit cells of zigzag blue phosphorene nanoribbons we studied, whose both edges are passivated with -H, -O, and -OH functional groups, respectively. We label the width of ZBPNRs with the number of parallel zigzag chains. The ZBPNRs with $n$ chains is thus named as $n$-ZBPNRs. To evaluate the structural stability of ZBPNRs with different groups of -H, -O, and -OH, we calculated their binding energy $E_b$, defined as $-(E_t - E_{BP} - 2*E_a)/2$, where $E_t$ is the total energy of ZBPNRs passivated with groups, $E_{BP}$ is the energy of unpassivated ZBPNRs, and $E_a$ represents the energy of an isolated group. The results demonstrate that the ZBPNRs-O have the largest $E_b$ and are the most stable structures, while the ZBPNRs-H have the smallest one. For example, the $E_b$ of the 5-ZBPNR-H, 5-ZBPNR-O and 5-ZBPNR-OH is 4.3, 6.6 and 4.4 eV, respectively.

These -H, -O, and -OH functional groups obviously tune the electronic structures of ZBPNRs and arouse them displaying distinguishable band structures. For example, the ZBPNRs-H present a wide gap semiconductor property, and their indirect band gap (~2.0 eV) changes little with increasing the ribbon width. It is different from the edge-passivated ZPNRs, which have a direct bandgap (~1.4 eV) [36], [37]. When the both edges of ZBPNRs are passivated with -O functional group, they show a metallic characteristic with two nearly degenerate bands (*i.e.* the band **I** and **II** dominated by O atoms, see Fig. 1(e)) crossing the Fermi level ($E_F$). The ZBPNRs-O are metallic and same to the edge-oxidized ZPNRs. It is because oxygen atoms form weak unsaturated bonds with the $p_z$ orbital of P atoms and bring some new edge states within the band gap. These edge states of ZBPNRs-O present around the $E_F$ within the band gap, closing up the band gap of ZBPNRs-O [38]. Moreover, when the both edges of ZBPNRs are passivated with -OH group, they show a similar semiconductor property to the ZBPNRs-H, but with a direct band gap (~2 eV) which is independent on the ribbons width. Monolayer TMDs were found to be a direct band gap semiconductor (1.0−2.0 eV) and are thus favorite for the logic transistor applications. The ZBPNRs-H and ZBPNRs-OH have the comparable band gaps and may also have potential applications in the field of logic transistors [39], [40].

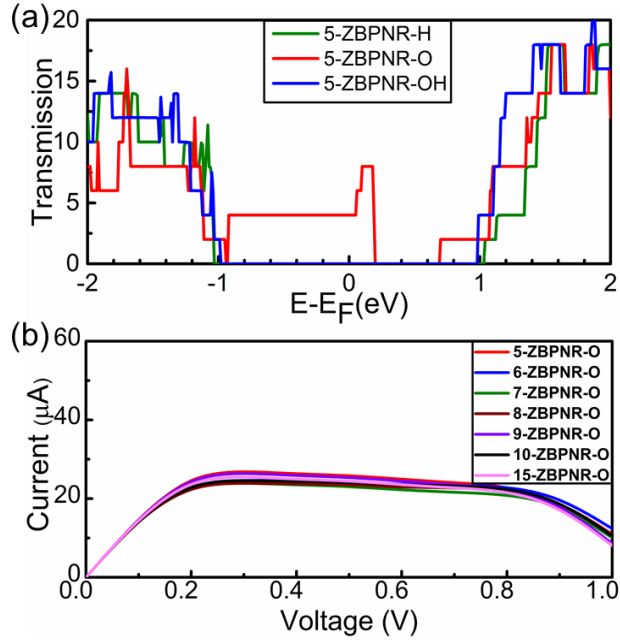

Fig. 3. (a) The transmission spectra of 5-ZBPNRs with different functional groups at the bias of 0 V. (b) The *I-V* curves of the ZBPNRs-O with different widths.

It can obviously tune the electronic structures of ZBPNRs through passivating the both edges with such common functional groups. To further understand the effects of functional groups modifying the edges, we investigate the electronic transport properties of ZBPNRs. As shown in Fig. 2, we build a two-probe device model of the zigzag blue phosphorene nanoribbons. It is composed of three parts: the central scattering region (C), which is about 25 Å in length, the left (L) and the right (R) electrodes. Both the left and the right electrodes are consisted of three repeated unit cells and applied periodic boundary condition.

According to the first-principles calculation results, the semiconductor ZBPNRs-H and ZBPNRs-OH indeed present a current forbidden behavior. While, the metallic ZBPNRs-O are conductive. As shown in Fig. 3(a), the electron transmission spectra of the 5-ZBPNR-H and 5-ZBPNR-OH are similar, and they both present a larger gap near the $E_F$, which suggests that little electric currents can propagate through these nanoribbons under the limited bias voltages. However, the 5-ZBPNR-O has a larger transmission coefficient near the $E_F$, displaying good electroconductibility

(see Fig. 3(a)). And its transmission spectrum displays a remarkable quantized characteristic. It is mainly due to the distinctive band structures of the 5-ZBPNR-O (see Fig. 1(e)).

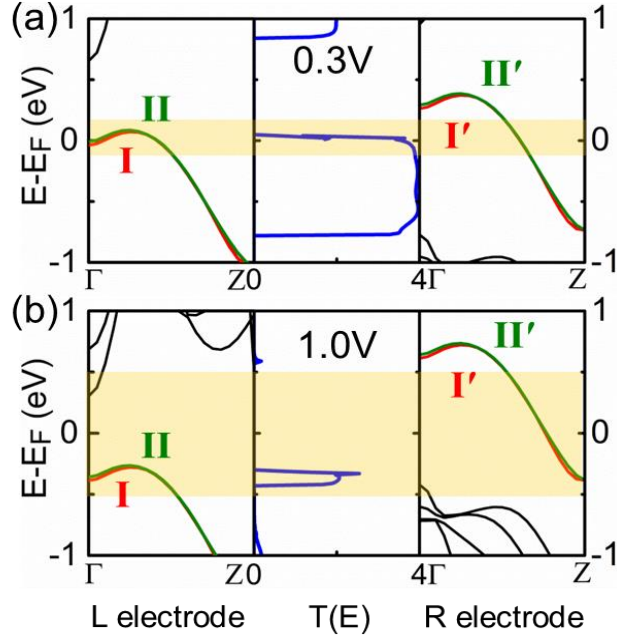

Fig. 4. The band structures for the L and R electrodes, and transmission spectra under the bias of 0.3 V (a) and 1.0 V (b) for the 5-ZBPNR-O. The shadows denote the bias window.

The current-voltage characteristic is one of the main parameters to evaluate the performance of low-dimensional nano-devices. To obtain the *I-V* curves of these ZBPNRs-based materials, we applied the external bias voltage $V_b$ ($V_b = V_L - V_R$, and $V_{L/R}$ is the bias voltage applied on the left/right electrode) to the two-probe device model. We use the Landauer–Büttiker formula [41] to describe the generated electric current *I*,

$$I(V_b) = \frac{2e}{h} \int_{-\infty}^{\infty} T(E, V_b)[f_L(E - \mu_L) - f_R(E - \mu_R)] dE \quad (1)$$

More details of the calculation method can be found in the previous reports [30]–[32].

In order to further understand the electronic transport properties of ZBPNRs-O under limited bias voltages, the *I–V* curves of the ZBPNRs-O with various ribbon widths are calculated and depicted in Fig. 3(b). One can see that the ZBPNRs-O show the similar electronic transport characteristics, and their *I–V* curves overlap with each other, showing little dependency on the ribbon width. What's more, the *I–V* curves display a NDR effect, and the currents reach the maximum value at the bias of 0.3 V. It suggests that the ZBPNRs-O can be used as a candidate of NDR nano-devices, and have the potential applications in nano-electronics, *e.g.*, high-frequency oscillators, multipliers, logic, and analog-to-digital converters.

In general, for one-dimensional nanoribbons, their electronic structures (*i.e.*, band structures) dominant the electronic transport behaviors. The NDR effect of the ZBPNRs-O can be understood most directly by the transmission spectra and band structures of the left and right electrodes under the biases of 0.3 and 1.0 V (see Fig. 4). The bands of the L and R electrodes shift downward and upward under a positive applied bias. The L and R electrode regions have an equilibrium electron distribution with their chemical potentials $\mu_L$ and $\mu_R$, related to the applied bias voltage, $\mu_L - \mu_R = eV_b$. The electrons with energies in the bias window (BW) $\mu_L \leq E \leq \mu_R$, give rise to a steady state electric current from the left electrode to the right one. With the increase of bias window, the overlapping area between the **I/II** and **I′/II′** bands of the electrodes is reduced, and the transmission coefficient decreases with the decrease of band overlaps. As shown in Fig. 4(a), the energy band of the left electrode has the most overlap with the right electrode at 0.3 V. The electron transmission coefficient is about 4, and the current reaches the peak value (about 25μA). However, when the bias voltage is increased to 1.0 V, the overlaps between the left and the right electrode bands decrease a little (see Fig. 4(b)), which results in a significant decrease in the electron transmission coefficient in the BW. Correspondingly, the electric current is reduced to about 9 μA. And then the NDR phenomenon is appeared subsequently (see Fig. 3(b)).

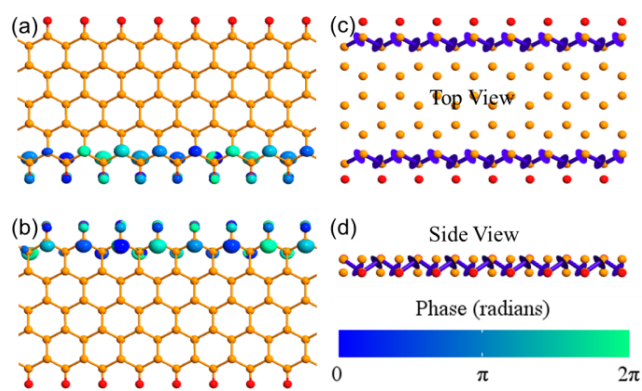

Fig. 5. (a) and (b) show the transmission eigenstates of the 5-ZBPNR-O under the bias of 0.3 V. (c) and (d) show the top and side views of the transmission pathways under 0.3 V bias, respectively. The isovalue of transmission eigenstates is set to 0.44 Å$^{(-3/2)}$.

To understand the intuitive electronic transmission mechanism of the ZBPNRs-O, we further analyzed the transmission eigenstates and transmission pathways (*i.e.* local currents) [42] at the transmission peak (E = 0 eV) under the bias of 0.3 V, as shown in Fig. 5 and 6. For the *odd* chains, taking the 5-ZBPNR-O as an example, there are two degenerate transmission eigenstates (see Fig. 5(a) and (b)), which are mainly located at one side of the nanoribbons. And the transmission eigenstates are mostly contributed by the 3*p* orbitals of phosphorus atoms.

Transmission pathway is an analysis option which splits the transmission coefficient into local bond contributions, $T_{ij}$. If the transmission pathways of the system is divided into two parts (A, B), the total transmission coefficient is obtained through the sum of pathways A and B

$$T(E) = \sum_{i \in A, j \in B} T_{ij}(E) \tag{2}$$

The arrows refer to the direction of electric current. As shown in Fig. 5(c) and (d), for the 5-ZBPNRs-O, there are two parallel transmission pathways, which are located at the both edges of nanoribbons. Moreover, one can see that the local currents are completely composed of the P-P bond current type (*i.e.*, electric current *via* the P-P chemical bond). It is noted that, the outside oxygen atoms give some contributions to the transmission eigenstates. However, they are too localized and give little contribution to the electron transmission.

Interestingly, the transmission eigenstates of the ZBPNRs-O show an *odd-even* effect. The transport eigenstates of the ZBPNRs-O with *odd* chains are only distributed on one side of the nanoribbons. For the ZBPNRs-O with *even* chains, however, taking the 6-ZBPNR-O as an example, it has two degenerate transmission eigenstates (see Fig. 6(a) and (b)), which are mainly located at the phosphorous atomic chains of both edges of nanoribbons. Moreover, the transmission channels of ZBPNRs-O with *even* chains has two transport pathways on the phosphorous atomic chains of both edges of nanoribbons. And they are symmetrical relative to the Z axis rather than parallel. The electric currents flowing through the 6-ZBPNR-O are mostly composed of the P-P bond current.

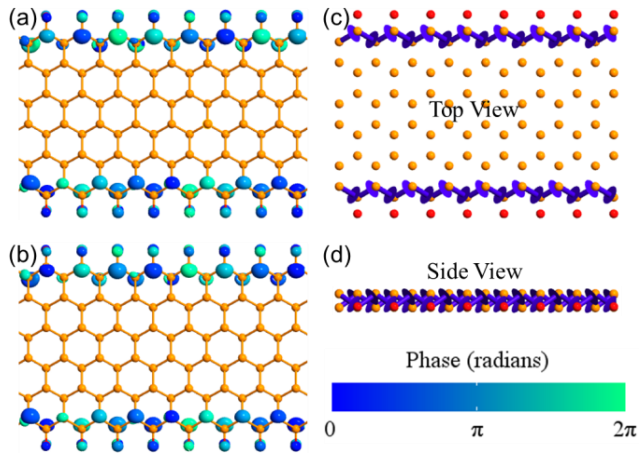

Fig. 6. (a) and (b) show the transmission eigenstates of the 6-ZBPNR-O under the bias of 0.3 V. (c) and (d) show the top and side views of the transmission pathways at 0.3 V bias, respectively. The isovalue of transmission eigenstates is set to 0.31 Å$^{(-3/2)}$.

## IV. CONCLUSIONS

In this paper, the electronic transport properties of zigzag blue phosphorene nanoribbons are investigated using first-principles method. Our calculations results demonstrate that, it can be tuned obviously through passivating the both edges of nanoribbons with some functional groups (*i.e.*, -H, -O, and -OH). The ZBPNRs-H present a wide gap semiconductor property, and their indirect band gap changes little with increasing the ribbon width. While, the ZBPNRs-O display the metallic characteristics, and their *I-V* curves show an interesting NDR effect, which are independent on their ribbon widths. The electric currents through ZBPNRs-O are mainly flowing along the both outside zigzag phosphorus chains *via* the way of P-P bond current. The *odd* and *even* ZBPNRs-O have the parallel and symmetrical electron transmission pathways, respectively. Moreover, the ZBPNRs-OH also display a semiconductor property but with a direct band gap. Our results suggest that modifying the edges with some functional groups is an effective method to tune the electronic structures and transport properties of zigzag blue phosphorene nanoribbons. And the proposed ZBPNRs-O could become the candidates of the nanodevices with the NDR effect.

REFERENCES

[1] K. S. Novoselov, A. K. Geim, S. V. Morozov, D. Jiang, Y. Zhang, and S. V. Dubonos, "Electric field effect in atomically thin carbon films," *Science*, vol. 306, no. 5696, pp. 666-669, Oct. 2004, doi: 10.1126/science.1102896.

[2] K. S. Novoselov, A. K. Geim, S. V. Morozov, D. Jiang, M. I. Katsnelson, I. V. Grigorieva, S. V. Dubonos, and A. A. Firsov, "Two-dimensional gas of massless Dirac fermions in graphene," *Nature*, vol. 438, pp. 197-200, Nov 2005, doi: 10.1038/nature04233.

[3] Z. Q. Fan, X. Q. Deng, G. P. Tang, C. H. Yang, L. Sun, and H. L. Zhu, "Effect of electrode twisting on electronic transport properties of atomic carbon wires," *Carbon*, vol. 98, pp. 179-186, May 2016, doi: 10.1016/j.carbon.2015.11.011.

[4] Y. P. An and Z. Q. Yang, "Abnormal electronic transport and negative differential resistance of graphene nanoribbons with defects," *Appl. Phys. Lett.*, vol. 99, no. 19, p. 192102, 2011, doi: 10.1063/1.3660228.

[5] Y. P. An, X. Y. Wei, and Z. Q. Yang, "Improving electronic transport of zigzag graphene nanoribbons by ordered doping of B or N atoms," *Phys. Chem. Chem. Phys.*, Vol. 14, no. 45, pp. 15802-15806, 2012, doi: 10.1039/c2cp42123b.

[6] Y. P. An, K. D. Wang, Z. Q. Yang, Z. Y. Liu, G. R. Jia, Z. Y. Jiao, T. X. Wang, and G. L. Xu, "Negative differential resistance and rectification effects in step-like graphene nanoribbons," *Organic Electronics*, vol. 17, pp. 262-269, Feb 2015, doi: 10.1016/j.orgel.2014.12.013.

[7] H. Zeng, C. Zhi, Z. Zhang, X. Wei, X. Wang, W. Guo, Y. Bando, and D. Golberg, "White Graphenes: Boron Nitride Nanoribbons via Boron Nitride Nanotube Unwrapping," *Nano Lett.*, vol. 10, no. 12, pp. 5049-5055, Dec 2010, doi: 10.1021/nl103251m.

[8] Y. P. An, K. D. Wang, G. R. Jia, T. X. Wang, Z. Y. Jiao, Z. M. Fu, X. L. Chu, G. L. Xu, and C. L. Yang, "Intrinsic negative differential resistance characteristics in zigzag boron nitride nanoribbons," *RSC Adv.*, vol. 4, no. 87, pp. 46934-46939, 2014, doi: 10.1039/c4ra08257e.

[9] Y. P. An, M. J. Zhang, D. P. Wu, T. X. Wang, Z. Y. Jiao, C. X. Xia, Z. M. Fu, and K. Wang, "The rectifying and negative differential resistance effects in graphene/h-BN nanoribbon heterojunctions," *Phys. Chem. Chem. Phys.*, vol. 18, no. 40, pp. 27976-27980, 2016, doi: 10.1039/c6cp05912k.

[10] H. Sahin, S. Cahangirov, M. Topsakal, E. Bekaroglu, E. Akturk, R. T. Senger, and S. Ciraci, "Monolayer honeycomb structures of group-IV elements and III-V binary compounds: First-principles calculations," *Phys. Rev. B*, vol. 80, no. 15, p. 155453, Oct 2009, doi: 10.1103/PhysRevB.80.155453.

[11] Y. P. An, M. J. Zhang, L. P. Chen, C. X. Xia, T. X. Wang, Z. M. Fu, Z. Y. Jiao, and G. L. Xu, "Spin-dependent electronic transport properties of zigzag silicon carbon nanoribbon," *RSC Adv.*, vol. 5, no. 129, pp. 107136-107141, 2015, doi: 10.1039/c5ra24276b.

[12] K. F. Mak, C. Lee, J. Hone, J. Shan, and T. F. Heinz, "Atomically Thin MoS2: A New Direct-Gap Semiconductor," *Phys. Rev. Lett.*, vol. 105, no. 13, p. 136805, Sept 2010, doi: 10.1103/PhysRevLett.105.136805.

[13] H. R. Gutiérrez, N. Perea-López, A. L. Elías, A. Berkdemir, B. Wang, R. T. Lv, F. López-Urías, V. H. Crespi, H. Terrones, and M. Terrones, "Extraordinary Room-Temperature Photoluminescence in Triangular WS2 Monolayers," *Nano*

*Lett.*, vol. 13, no. 8, pp. 3447-3454, Aug 2013, doi: 10.1021/nl3026357.

[14] Y. P. An, M. J. Zhang, H. X. Da, Z. M. Fu, Z. Y. Jiao, and Z. Y. Liu, "Width and defect effects on the electronic transport of zigzag MoS2 nanoribbons," *J. Phys. D: Appl. Phys.*, vol. 49, no. 24, p. 245304, 2016, doi: 10.1088/0022-3727/49/24/245304.

[15] Y. P. An, M. J. Zhang, D. P. Wu, Z. M. Fu, and K. Wang, "The electronic transport properties of transitionmetal dichalcogenide lateral heterojunctions," *J. Mater. Chem. C.*, vol. 4, no. 46, pp. 10962-10966, 2016, doi: 10.1039/C6TC04327E.

[16] L. K. Li, Y. J. Yu, G. J. Ye, Q. Q. Ge, X. D. Ou, H. Wu, D. L. Feng, X. H. Chen, and Y. B. Zhang, "Black phosphorus field-effect transistors," *Nat. Nanotechnol.*, vol. 9, pp. 372-377, May 2014, doi: 10.1038/NNANO.2014.35.

[17] J. L. Zhang, S. T. Zhao, C. Han, Z. Z. Wang, S. Zhong, S. Sun, R. Guo, X. Zhou, C. D. Gu, K. D. Yuan, Z. Y. Li, and W. Chen, "Epitaxial Growth of Single Layer Blue Phosphorus: A New Phase of Two-Dimensional Phosphorus," *Nano Lett.*, vol. 16, no. 8, pp. 4903-4908, Aug 2016, doi: 10.1021/acs.nanolett.6b01459.

[18] J. Xiao, M. Long, X. Zhang, J. Ouyang, H. Xu, and Y. Gao, "Theoretical predictions on the electronic structure and charge carrier mobility in 2D Phosphorus sheets," *Sci. Rep.*, vol. 5, p. 9961, Jun 2016, doi: 10.1038/srep09961.

[19] F. Li, X. H. Liu, Y. Wang, and Y. F. Li, "Germanium monosulfide monolayer: a novel two-dimensional semiconductor with a high carrier mobility," *J. Mater. Chem. C.*, vol. 4, no. 11, pp. 2155-2159, 2016, doi: 10.1039/c6tc00454g.

[20] M. J. Zhang, Y. P. An, Y. Q. Sun, D. P. Wu, X. N. Chen, T. X. Wang, G. L. Xu, and K. Wang, "The electronic transport properties of zigzag phosphorene-like MX (M = Ge/Sn, X = S/Se) nanostructures," *Phys. Chem. Chem. Phys.*, vol. 19, no. 26, pp. 17210-17215, 2017, doi: 10.1039/c7cp02201h.

[21] H. Liu, A. T. Neal, Z. Zhu, Z. Luo, X. F. Xu, D. Tomanek, and P. D. Ye, "Phosphorene: An Unexplored 2D Semiconductor with a High Hole Mobility," *ACS Nano*, vol. 8, no. 4, pp. 4033-4041, Apr 2014, doi: 10.1021/nn501226z.

[22] S. P. Koenig, R. A. Doganov, H. Schmidt, A. H. C. Neto, and B. Özyilmaz,

"Electric field effect in ultrathin black phosphorus," *Appl. Phys. Lett.*, vol. 104, no. 10, p. 103106, 2014, doi: 10.1063/1.4868132.

[23] F. Xia, H. Wang, and Y. Jia, "Rediscovering black phosphorus as an anisotropic layered material for optoelectronics and electronics," *Nature Commun*, vol. 5, p. 4458, Apr 2014, doi: 10.1038/ncomms5458.

[24] V. Tran, R. Soklaski, Y. Liang, and L. Yang, "Layer-controlled band gap and anisotropic excitons in few-layer black phosphorus," *Phys. Rev. B*, vol. 89, no. 23, p. 235319, Jun 2014, doi: 10.1103/PhysRevB.89.235319.

[25] D. Çakir, H. Sahin, and F. M. Peeters, "Tuning of the electronic and optical properties of single-layer black phosphorus by strain," *Phys. Rev. B*, vol. 90, no. 20, p. 205421, 2014, doi: 10.1103/PhysRevB.90.205421.

[26] F. Xie, Z. Q. Fan, X. J. Zhang, J. P. Liu, H. Y. Wang, K. Liu, J. H. Yu, and M. Q. Long, "Tuning of the electronic and transport properties of phosphorene nanoribbons by edge types and edge defects, "*Organic Electronics*, vol. 42, pp. 21-27, Mar 2017, doi: 10.1016/j.orgel.2016.12.020.

[27] Z. Zhu and David Tománek, "Semiconducting Layered Blue Phosphorus: A Computational Study," *Phys. Rev. Lett.*, vol. 112, no. 17, p. 176802, May 2014, doi: 10.1103/PhysRevLett.112.176802.

[28] J. L. Zhang, S. T. Zhao, C. Han, Z. Z. Wang, S. Zhong, S. Sun, R. Guo, X. Zhou, C. D. Gu, K. D. Yuan, Z. Y. Li, and W. Chen, "Epitaxial Growth of Single Layer Blue Phosphorus: A New Phase of Two-Dimensional Phosphorus, " *Nano Lett.*, vol. 16, no. 8, pp. 4903-4908, Aug 2016, doi: 10.1021/acs.nanolett.6b01459.

[29] J. Xiao, M. Q. Long, C. S. Deng, J. He, L. L. Cui, and H. Xu, "Electronic Structures and Carrier Mobilities of Blue Phosphorus Nanoribbons and Nanotubes: A First-Principles Study," *J. Phys. Chem. C.*, vol. 120, no. 8, pp.4638-4646, Mar 2016, doi: 10.1021/acs.jpcc.5b12112.

[30] M. Brandbyge, J. L. Mozos, P. Ordejon, J. Taylor, and K. Stokbro, "Density-functional method for nonequilibrium electron transport," *Phys. Rev. B*, vol. 65, no. 16, p. 165401, Mar 2002, doi: 10.1103/PhysRevB.65.165401.

[31] J. Taylor, H. Guo, and J. Wang, "Ab initio modeling of open systems: Charge

transfer, electron conduction, and molecular switching of a C60 device," *Phys. Rev. B*, vol. 63, no. 12, p. 121104, Jun 2001, doi: 10.1103/PhysRevB.63.121104.

[32] J. M. Soler, E. Artacho, J. D. Gale, A. García, J. Junquera, P. Ordejón, and D. Sánchez-Portal, "The SIESTA method for ab initio order-N materials simulation, "*J. Phys.: Condens. Matter*, vol. 14, no. 11, p. 2745, 2002, doi: 10.1088/0953-8984/14/11/302.

[33] See http://quantumwise.com/ for Atomistix ToolKit

[34] J. P. Perdew, J. A. Chevary, S. H. Vosko, K. A. Jackson, M. R. Pederson, D. J. Singh, and C. Fiolhais, "Atoms, molecules, solids, and surfaces: Applications of the generalized gradient approximation for exchange and correlation," *Phys. Rev. B*, vol. 46, no. 11, pp. 6671-6687, Sept 1992, doi: 10.1103/PhysRevB.46.6671.

[35] J. P. Perdew, K. Burke, and M. Ernzerhof, "Generalized Gradient Approximation Made Simple," *Phys. Rev. Lett*, vol. 77, no. 18, pp. 3865-3868, Oct 1996, doi: 10.1103/PhysRevLett.77.3865.

[36] W. F. Li, G. Zhang, and Y. W. Zhang, "Electronic Properties of Edge-Hydrogenated Phosphorene Nanoribbons: A First-Principles Study," *J. Phys. Chem. C*, vol. 118, no. 38, pp. 22368-22372, Sept 2014, doi: 10.1021/jp506996a.

[37] H. Y. Guo, N. Lu, J. Dai, X. J. Wu, and X. C. Zeng, "Phosphorene Nanoribbons, Phosphorus Nanotubes, and van der Waals Multilayers," *J. Phys. Chem. C*, vol. 118, no. 25, pp. 14051-14059, Jun 2014, doi: 10.1021/jp505257g.

[38] X. H. Peng, Q. Wei, and Andrew Copple, "Edge effects on the electronic properties of phosphorene nanoribbons," *J. Appl. Phys*., vol. 116, no, 14, p. 144301, 2014, doi: 10.1063/1.4897461. [39] Z. Q. Fan, X. W. Jiang, J. Z. Chen, and J. W. Luo, "Improving Performances of In-Plane Transition-Metal Dichalcogenide Schottky Barrier Field-Effect Transistors," *ACS Appl. Mater. Interfaces*, vol. 10, no. 22, pp. 19271-19277, May 2018, doi: 10.1021/acsami.8b04860.

[40] Z. Q. Fan, X. W. Jiang, J. W. Luo, Y. J. Li , R. Huang, S. S. Li, and W. W. Lin, "In-plane Schottky-barrier field-effect transistors based on 1T/2H

heterojunctions of transition-metal dichalcogenides," *Phys. Rev. B*, vol. 96, no. 16, pp. 16, 5402-165407, Sept 2017, doi: 10.1103/PhysRevB.96.165402.

[41] M. Büttiker, Y. Imry, R. Landauer, and S. Pinhas, "Generalized many-channel conductance formula with application to small rings," *Phys. Rev. B: Condens. Matter Mater. Phys.*, vol. 31, no. 10, pp. 6207-6215, May 1985, doi: 10.1103/PhysRevB.31.6207.

[42] G. C. Solomon, C. Herrmann, T. Hansen, V. Mujica, and M. A. Ratner, "Exploring local currents in molecular junctions," *Nat. Chem.*, vol. 2, pp. 223-228, Feb 2010, doi: 10.1038/NCHEM.546.